\documentclass[12pt]{article}
\usepackage{graphicx}

\begin{document}

\section* {\bf CAN QUANTUM MECHANICS BE CLEARED FROM CONCEPTUAL
 DIFFICULTIES?}

\medskip\bigskip

\begin{center}
{\large {\bf Volodymyr Krasnoholovets}}
\end{center}

\medskip

\begin{center}
{Institute of Physics, National Academy of Sciences \\ Prospect
Nauky 46, UA-03028 Ky\"{\i}v, Ukraine \\  http://inerton.cjb.net}
\end{center}

\medskip

\hspace{11 cm} {20 July 2002}

\begin{center}
{\bf Abstract}
\end{center}

The major conceptual difficulties of quantum mechanics are
analyzed. They are: the notion "wave-particle", the probabilistic
interpretation of the Schr\"odinger wave $\psi$-function and hence
the probability amplitude and its phase, long-range action,
Heisenberg's uncertainty principle, etc. The probabilistic
formalism is developed in the phase space, but not in the real
one. Elimination of the difficulties is likely if we are able to
develop quantum mechanics in the real space. Such a theory in fact
can be constructed, however, it should proceed from deepest first
principles starting from the notion of a 4D space-time, the notion
of a massive particle in the space, the principles of the motion
of a particle, etc. The theory should be characterized by
short-range action that automatically means the introduction of a
quantum mechanical force. It is shown that the aforementioned
force makes it evident and, moreover, is able to appear on the
macroscopic scale. A simple experiment, the express test, which in
fact proves the macroscopic manifestation of quantum mechanical
force, is proposed for the demonstration in the quantum
curriculum.

\bigskip
\bigskip

{\bf  Key words:} \ \  quantum mechanics, space, matter waves,
inertons
\\

{\bf PACS:} \ \  03.65.Bz Foundations, theory of measurement,
miscellaneous theories; \ 03.65.w Quantum mechanics; \ 03.75.-b
Matter waves; \ 14.80.-j Other particles (including hypothetical)

\newpage

\begin{flushright} \footnotesize{Today we shall soberly
look at what we expect from tomorrow.   \ \ \ \  \ \ \ \ \ \ \ \ \
\ \ \ \ \ \ \ \ \ \ \ \ \ \ \ \ \ \ \ \ \ \ \ \ \ \ \ \ \
\
\
\
\
\
\
\
\hspace{5cm}   E. P. Wigner, {\it Symmetries and Reflections}\ \
}
\end{flushright}

\section{Introduction}

\hspace*{\parindent} Eugene Wigner was keenly interested in
mathematical problems of interpretation of quantum mechanical laws
and, in particular, he actively studied the problems of measuring.
His studies arrived him at the formulation of the following
problems of orthodox quantum mechanics, i.e. its difficulties [1]:
(i) the vector of state "the object plus the instrument" cannot be
distinguished from the mixture of states; (ii) a possibility of
the measuring only state $\sigma^{(\nu)}$ of a quantum system can
be measured only in the case when the instrument is a very large
system; (iii) quantum laws allow one to obtain only probabilistic
correlations between results of several consequence observations
of the quantum system. He particularly emphasized the importance
of a macroscopic instrument in the problem of measuring: just the
instrument should reduce the wave package (or sometimes the vector
of state). While on the subject of hidden variables, Wigner [2]
noted that we could not obtain directly their magnitudes. He
noticed that the measuring the distribution of magnitudes of
hidden variables still remained undetermined, or fuzzy.

Nowadays, however, researchers concern on some other aspects of
the foundations of quantum mechanics, which have been revealed in
the course of recent studies associated with the phenomenon of
entanglement states. One of the main problems which is examined at
present is a possibility for nonlocality of quantum theory. Bell
[3] introduced some new aspects to the problem of completeness of
quantum theory. He formulated a locality requirement introducing
some additional variables, so called local hidden variables. Such
a study initiated a long series of thought experiments, which then
resulted in some actual experiments (however, the experiments
involved photons as quantum entities, which, strictly speaking,
are entities of quantum electrodynamics, but not quantum
mechanics). In particular, we should mention here research by
Stapp [4,5] who has carried out a detailed analysis of both
theoretical and experimental results, which touch questions like
these: Is quantum theory local or nonlocal? and Is nonlocality is
real? Stapp has adduced many arguments for this or that point of
views and specifically noted that quantum theory is still
formulated as an indeterministic theory.

Indeterminism is a very important starting point of modern quantum
theory. However, would we are based on the other assumption?..
This question indeed is very interesting as the searching for the
answer could clarify the major fundamental physical notions and
culminate in the discovery of a series of new links between them,
which so far were still hidden from researchers including those
who follow Louis de Broglie and David Bohm, pioneers of
determinism.

\medskip

\section{White gaps in quantum concept}

\hspace*{\parindent} {\bf \underline{1.}} \ First of all let us
take a good look at the term "conceptual difficulties". The term
implies that the doctrine under consideration features strong
discrepancies between characteristics, which it describes, and
methods, which the doctrine uses. In the case of quantum
mechanics, the situation is dramatized by the fact that one more
characteristic should be ascribed to a canonical particle, namely,
the particle as such is transformed to a certain "particle-wave".
And this is the first conceptual difficulty of quantum mechanics!
Indeed, how can one understand the particle-wave? In 1924 de
Broglie, when wrote his remarkable relationships

\begin{equation}    E=h\nu \quad \ \ \
{\rm and} \ \ \ \quad \   \lambda =h/p,  \label{1}
\end{equation}

\medskip \noindent assumed that some real wave was connected with
the moving particle and that just this wave guided the particle.
In expressions (1) parameters $E$ and $p$ (the energy and the
momentum) belonged to the particle, but the frequency $\nu$ and
the wavelength $\lambda$ were characteristics of a wave that
should accompany the particle at its motion in the real space.
Especially as relationships (1) enable one to derive the
Schr\"odinger equation [6].

\underline{{\bf Corollary 1.}} De Broglie's transparent idea that
a moving particle is accompanied by an actual wave did not receive
any further development.

\vspace{4mm}

{\bf \underline{2.}} \  The Schr\"odinger equation written in 1925
was successfully applied to the calculation of energies of
equilibrium states of an electron in the Coulomb potential of a
proton, which practically coincided with the experimentally
measured spectrum of the hydrogen atom. Such an excellent
correspondence between the prediction of the theory and the
experimental results gave immediate impetus to the construction of
the probabilistic formalism of quantum mechanics. Born's and
Heisenberg's abstract formalism replaced de Broglie's common
sense. Thus, Born's probabilistic interpretation of the
Schr\"odinger wave $\psi$-function rejected any conceivable
physical content from the $\psi$. Nowadays a quantum system is
described by the probability amplitude $|\psi|^2$ and its phase
$\phi$ that includes information on the energy, momentum and
coordinate of the particle and it is also implied that $\phi$
involves information on the wave characteristics of the particle,
some frequency $\nu$ and wavelength $\lambda$.

\underline{{\bf Corollary 2.}} In the modern interpretation, the
wave $\psi$-function is quite abstract. However, it is believed
that at the measuring process the abstract wave function collapses
to a measurable actual point particle.

\vspace{4mm}

{\bf \underline{3.}} \ At the same time, the existence of the
actual wave properties in particles, i.e. the matter waves,
received empirical confirmation in the diffraction experiments.
Therefore, particles in fact possess wave properties and this
automatically implies that the pure probabilistic interpretation
of the $\psi$-function is not complete.

Recently Briner et al. [7] has published an experimental work
entitled "Looking at Electronic Wave Functions on Metal Surfaces",
in which they demonstrate the colored spherical and elliptical
figures, which the authors called "the images of $\psi$ wave
functions of electrons". Virtually they gave the evidence that the
electron is not a point-like object, though the high energy
physics asserts that it is a point object with the size no larger
than $10^{-17}$ cm. Thus they fixed an actual perturbation of the
space around an electron in the metal! Thereby, the authors
subconsciously rose against the probabilistic interpretation of
the $\psi$ wave accepted by the Copenhagen School concept and,
moreover, they practically proved the fallaciousness of the
statement of the concept.

\underline{{\bf Corollary 3.}} Experimental data point to the fact
that the wave $\psi$-function is not abstract but a measurable
matter.

\vspace{4mm}

{\bf \underline{4.}} \ Furthermore, the Schr\"odinger and Dirac
formalisms say nothing about true trajectories of the quantum
system studied that is a direct consequence of the probabilistic
approach to the description of quantum phenomena. Of course, one
could use Feynman diagrams for any entity, with their point-like
particles and photons, all having some absolute position and
momentum. However, we cannot get the true path. Instead we must
draw infinitely many Feynman diagrams and then calculate the
Feynman's path integrals, which make it possible to find out only
the most verisimilar trajectory of the quantum system.

Once again, this is because of the fact that conventional quantum
mechanics is developed in the phase space, but not in the real
one. Indeed, can one clarify the duality of a "particle-wave" in
the real space where only a particle and a wave can separately be
determined? The same is noted by Ligare and Olivery [8]: "it is
not always clear which aspects of classical wave behaviour are
related in a fully quantum-mechanical treatment, or where to draw
the line between wave-like aspects and particle-like aspects and
how to justify the division".

When we talk about the real space we imply a 3D space or a 4D
space-time, in which one can assign exact position, velocity and
momentum to an object at any time. A wave can also simply be given
in a 3D space or 4D space-time, but in this case the space should
possess clear defined condensed matter properties.

\underline{{\bf Corollary 4.}} If we wish to understand the
"particle-wave", we must turn to the consideration of quantum
mechanics in a space filled with a subquantum medium that was
first pointed out by de Broglie (see e.g. Ref. [9]).

\vspace{4mm}

{\bf \underline{5.}} \ Next negative aspect is that the
probabilistic formalism severe suffers from long-range action. By
conventional quantum mechanics, particles can interact
simultaneously even if they are spaced at any quantity of
kilometers, Ehrenfest [10]. Long-range action of quantum mechanics
was also emphasized by Pauli [11]; in particular, he noted that
quantum mechanics bears up against a hypothetical basis that the
speed of the interaction in the quantum mechanical range
$c=\infty$ and that the gravitational interaction is negligible,
the constant of gravitational interaction $G=0$.

   For instance, let us turn to the problem of hydrogen atom, a
typical example of long-range action in quantum mechanics. The
radial part of the Schr\"odinger equation written for a particle
in a spherically symmetric electrostatic potential $V(r)$ has the
form (see, e.g. Schiff [12])

  $$
-\frac{\hbar^2}{2m} \frac {{\rm d}^2 \chi}{{\rm d} r^2} + \Big[
V(r) + \frac {l(l+1) \hbar^2}{2mr^2} \Big] \chi = E \chi
 \eqno(2)
  $$

\medskip\noindent where $\chi(r)$ is the radial wave function.
The second term in the square brackets is stipulated by the
potential energy associated with the moment of momentum of the
particle. The potential energy

  $$
V(r) + \frac {l(l+1) \hbar^2}{2mr^2}   \eqno(3)
  $$
\medskip\noindent
ensures the stability of the particle orbit. In the case of the
hydrogen atom the potential $V(r) = e^2/(4\pi \epsilon_0{\kern
1pt} r)$ and the equation of related motion of an electron and
proton has the form similar to Eq. (2).

However, it should be noted that the Schr\"odinger {\it quantum}
equation (2) includes the potentials $V(r)$ written in pure {\it
classical} terms, much as in the problem of Newton gravity! $V(r)$
is a usual classical presentation of the motionless charge and the
electromagnetic field that surrounds it. The mass $m$ that enters
into {\it quantum} equation (2) is also a pure {\it classical}
parameter. Hence even the most comprehensive quantum mechanical
description of the quantum system studied is only a
quasi-classical pattern.

\underline{{\bf Corollary 5.}} \ If we remain devotees of orthodox
quantum mechanics, the fundamentals will be kept in the shade of
its statistical conformities.

\vspace{4mm}

{\bf \underline{6.}} \ Although there are Heisenberg's
uncertainties for the coordinate and momentum and the energy and
time of a particle,

  $$
\Delta x {\kern 1pt}\Delta p \geq \hbar; \ \ \ \ \ \Delta E {\kern
1pt} \Delta t \geq \hbar,    \eqno(4)
  $$

\medskip\noindent we are not able to write any similar relation
for the particle mass, which should also be fuzzy in a
undetermined volume, the same as the particle itself (the mass
must follow the particle!!), as the probabilistic formalism
prescribes.

De Broglie [13,14] studied this problem and came to the conclusion
that the dynamics of particles had the characteristics of the
dynamics of the particles with a variable proper mass. He was the
first to indicate that the corpuscle dynamics was the basis for
the wave mechanics.  With the variational principle, he obtained
and studied the equations of motion of a massive point reasoning
from the typical Lagrangian
  $$
L = -M_0c^{\kern 1pt 2}\surd\overline{1-v^2/c^{\kern 1pt 2}}
\eqno(5)
  $$
in which the velocity $v$ of the point and the velocity of light
$c$ were constant along a path. De Broglie's pioneer research
allows one to suggest that a real wave, which indeed has to
accompany the moving particle, must complement the deficient value
of the momentum and the energy of the particle. Then, say, we know
the momentum and the energy, but have uncertainties in coordinate
and time. If we assume the existence of an actual wave that
travels in the space together with the particle, we can readily
propose that the particle is entrained by the said wave and,
therefore, position and time of the particle become in fact
undetermined in a concrete point as they become functions of the
traveling wave.

\underline{{\bf Corollary 6.}} \ Heisenberg's uncertainty is a
direct consequence of the probabilistic approach to quantum
phenomena when only one of two subsystems is taken into account,
namely, we treat the behaviour only a particle, but totally ignore
the real wave, which accompany the particle.

\vspace{4mm}

{\bf \underline{7.}} \ All correct theories should be Lorentz
invariant, i.e. they and Einstein's special relativity should
agree (see, e.g. Ref. [15]). Nevertheless, the Schr\"odinger
equation is not Lorentz invariant but it perfectly describes
quantum phenomena and we  trust wholly the results derived from
the equation. How is it possible?

It seems that the disagreement between the strong theoretical
conclusion and the experimental veracity is hidden in the
statistical approach to the Schr\"odinger formalism. Indeed,
relationships (1) allow the derivation of the Schr\"odinger
equation as well [6], but what exactly do the relationships
describe? In recent papers by the author [16,17] the inner sense
of relationships (1) was studied in detail starting from an idea
that the physical reality represented a space net that came into
the interaction with a moving particle. This allowed the
derivation of the Schr\"odinger equation from deepest first
principles that in fact removed a very unpleasant conflict that so
far took place between nonrelativistic quantum mechanics and
special relativity: Unlike the traditional presentation, the
Schr\"odinger equation gained in paper [17] is Lorentz invariant
owing to the invariant time entered in the equation.

\underline{{\bf Corollary 7.}} \ The Schr\"odinger equation is
Lorentz invariant.

\vspace{4mm}

{\bf \underline{8.}} \ There is no correct determination of values
$E$ and $\nu$ in the expression $E = h \nu$ applied to a moving
canonical particle.  In one case $E=\frac 12 m_0{\kern 1pt}v^2$
(see, e.g. Schiff [12], p. 33 ), and in the other one
$E=m_0c^{\kern 1pt 2}(1-v^2/c^{\kern 1pt 2})^{-1/2}$ (see, e.g.
Schiff [12], p. 364). Which is true?

The problem has been studied by the author in paper [18], in which
the motion of a relativistic particle has been treated based on a
generalized lattice model of the real space. It has been shown
that if the moving particle interacts with the space, the feedback
governs the quantum system in question and the system undergoes
the phase transition when its velocity $v$ trends to $c$. In the
case $v << c$, an associated real wave, which guides the particle,
carries the particle's kinetic energy $E=\frac 12 M_0{\kern
1pt}v^2$; in the case $v \rightarrow c$ the associated wave
becomes closed inside of the range covered by the Compton
wavelength $\lambda_{\rm {\kern 1pt}} =h/mc$ of the particle and
hence the kinetic energy of the particle is given by the total
energy of the region, $E=m_0c^{\kern 1pt 2}(1-v^2/c^{\kern 1pt
2})^{-1/2}$.

\underline{{\bf Corollary 8.}} \ Allowance for the interaction of
the quantum system under consideration with the real space
clarifies difficult questions of quantum mechanics and, in
particular, gives the unambiguous answer to the question [18]:
What is nature of the phase transition, which occurs in the
quantum system, that turns us from the description of the system
based on the Schr\"odinger equation to that resting on the Dirac
one?

\vspace{4mm}

{\bf \underline{9.}} \  What is spin? It is one more mystery of
the microworld. In quantum mechanics spin is perceived to be a
certain inner property of canonical particles. Quantum field
theories define spin as an "inseparable and invariable property of
a particle" (see e.g. Ref. [15], p. 17). That is all.

As a rule the notion of spin of a particle is associated with an
intrinsic particle motion. Several tens of works have been devoted
to the spin problem. Major of them is reviewed in recent author's
papers [18,19]. Main ideas of the works quoted in Refs. [18,19]
are reduced to a moving particle that is surrounded by a wave, or
a small massless particle, or an ensemble of small massless
particles, which engage in a circular motion.

Of course, it seems quite reasonable to assume that spin in fact
reflects some kind of proper rotation of the particle. However,
canonical particles possess also electrodynamic properties and the
operation "rotor" is the principal characteristic of the particle
electromagnetic field. Therefore, the accord between quantum
electrodynamics and quantum mechanics of a particle requires the
abandonment of the idea of rotation with respect to the notion of
the particle spin. This means that we should associate the
rotational electromagnetic field generated by a canonical particle
with the particle's proper rotation of some sort.

Particle physics also cannot offer any reasonable answer to the
question on the problem of spin, as this branch of physics does
not deal with spatial images of particles which, nevertheless, are
the main subject of its study. If quantum mechanics considers
particles by means of their abstract $\psi$ functions, particle
physics treats the subject basing on all the more abstract notion
of fundamental symmetry.

In the author concept [16-19] particles are determined just as
spatial images (or objects) in the real space, which in fact makes
it possible to investigate the notion of spin in detail. In this
case along with an oscillating rectilinear motion, the particle
undergoes also some kind of an inner pulsation, like a drop. The
two possible orientations of pulsations either along the particle
velocity vector or diametrically opposite to it are associated
with the particle spin [18].

\underline{{\bf Corollary 9.}} \ The notion of the particle spin
can be determined only in the framework of quantum mechanics
constructed in the real space. Two possible own pulsations of the
particle in the real space are exhibited by two so-called spin-1/2
projections in the phase space. An integer-valued spin is the
property of a composite quantum system.

\vspace{4mm}

{\bf \underline{10.}} \ Dirac [20] considering links between
general relativity and quantum mechanics noted that although the
relativity posed the objections to an aether, quantum mechanics
practically removed them. This automatically means that a vacuum,
which is hazy something or nothing in all modern quantum theories
(quantum mechanics, quantum elecrodynamics, chromodynamics, etc.),
should be replaced by a concrete subquantum substrate. High energy
physics working on sub microscopic scales proposes some Higgs
condensate, which would be initial at the creation of the physical
world. Nonetheless, the Higgs condensate of models of grand
unification of interactions is not constructed in a real 4D
space-time and moreover, it does not give any idea in what way it
can manifest itself in quantum mechanics. We emphasize that it is
quantum mechanics that is the most reliable basis for all the
other quantum theories. Because of that any new quantum concept
should produce orthodox quantum mechanics as a limiting case of
the theory constructed. However, either quantum chromodynamics, or
some other contemporary theory (such as string theory) is not able
to mutate in the orthodox quantum mechanical formalism. Quantum
field theories and their derivatives suffer from undetermined
field variables $\varphi$, $\varphi^4$ and so on. Group methods
also isolate themselves from both the constitution of the space
and the direct measurement.

General relativity does not deal with any substrate, its major
subject is geometry. However, we should not forget that the
relativity separates the mass from the geometry, i.e. matter from
space. Nonetheless, if we assume that matter appears from the
space, which in turn is a substrate, we immediately arrive at the
conclusion that the matter should interact with such a space: the
space itself becomes material.

\underline{{\bf Corollary 10.}} \ A quantum theory constructed
with regard to the connection of a quantum system with the real
space in which the system is found will arrive us at very new
horizons in both the subatomic area (the strong and electroweak
interaction would be revised) and macroscopic one (the theory of
gravity would be developed starting from quantum mechanics as
well).

\vspace{4mm}

\section{Quantum mechanical force}

\hspace*{\parindent} As of now, quantum entanglement, the property
that allows two particles to behave as one, no matter how far
apart they are, has been much investigated. Experimentally, if we
measure the state of one particle, we instantly determine the
state of the other. Researchers probe the possibility to teleport
not just quantum states of photons, but also of more massive
particles. And it is anticipated that the phenomenon could one day
allow us to teleport objects by transferring their properties
instantly from one place to another.

In interesting theoretical work [21] interference and entanglement
have recently been studied in the frame of a so-called intrinsic
approach. As a rule classical states are prescribed to the phase
space, while quantum states, which are considered as linear
functions on the dynamical variables, assign to the vector space
formalism (the Hilbert space). Density matrices belong to
bilinears in the vectors. The authors [21] propose a generalized
method of restoring an entangled pure state (a purification of the
linear vector) from mixed states of the subsystems involved in
entanglement. The pure density matrix is then treated as a measure
of entanglement for the quantum system in question. In orthodox
quantum mechanics linear operators, which act on the Hilbert space
of states, are associated with observables. For instance, the
Wigner distribution $f(\bf x)$ is related to observables as well.
Nevertheless, although $f(\bf x)$ is defined in the Hilbert space
[21] its arguments belong to the phase space (note in the general
case the Wigner function is determined as $f({\bf x}, \; {\bf p},
\; t )$). So the phase space and the Hilbert one appear as
interconnected, though the Hilbert space describes strictly
stationary states of the quantum system in question.

From the view point of conceptual difficulties of quantum theory,
the entanglement represents some synthesis of nonlocality and
long-range action considered above. These two difficulties as well
as all the other ones are associated with the area of existence of
quantum mechanics, i.e. the phase space, in which orthodox quantum
mechanics is constructed. However, all the problems are remedied
by passing on to quantum mechanics derived on a deeper fundamental
basis, namely, the real space.

Such a theory, submicroscopic quantum mechanics, indeed has
recently been developed by the author (see self reviews [19]). It
is argued that a particle moving in the real space is surrounded
by a cloud of elementary excitations called "inertons", which
appear due to friction of particle-on-superparticles (where
superparticles are building blocks of the real space, see also
Refs. [22]). The particle along with its inerton cloud moves as a
typical real wave. Inertons, which accompany the particle,
represent a substructure of its matter waves and, because of that,
they are carriers of the particle inert properties.

Thus, just inertons teleport quantum states of one particles to
the other ones. The mass of inertons can easily be estimated [23].
Besides, the inerton and the photon are not fundamentally
different: It is an inerton that is an undercoat for the photon,
i.e. the photon is an inerton, which is supplemented by the
electromagnetic polarization [24]. We know that photons are
carriers of the electromagnetic interaction (or electromagnetic
force) between both quantum entities and macroscopic objects.
Therefore, since the photon is a complexly built quasi-particle
whose base is the inerton, we immediately get the conclusion that
inertons should be carriers of both the quantum mechanical
interaction (or quantum mechanical force), which acts between
quantum entities, and the pure mechanical interaction, which
occurs between macroscopic objects. In the last case inertons
manifest themselves as carriers of the force of inertia, which is
the major mechanical force. Indeed, everyone can recall that any
abrupt halt is attended with the further inexplicable jog. The
reason can simply be explained in the frame of the submicroscopic
concept: our own cloud of inertons continues to push us slightly
when we instantly pull up. In the case of a quantum system the
impact of inertons should intensify oscillations of a particle(s),
which in turn yields an option to guide the particle(s) to another
place, including the so-called tunneling. In other words, in the
quantum case a flow of inertons can be treated as a confinement
field, which exacts control over the system behaviour. Besides,
inertons having the energy and the momentum carry also local
deformations, which result in the induction of the gravitational
potential around a particle or a classical object [25].

One can put a question whether inertons, as carriers of the
quantum mechanical force, can be measured on the macroscopic
scale. Yes, they certainly can. Recent research has shown [26,27]
that Egyptian pyramids were functioning as peculiar plants, which
projected (or teleported?) the Earth by transferring its
properties to the pyramids. In other words, especially the Great
Pyramid of Giza was constructed (by modern estimates several tens
of thousand years ago [27]) as a transducer that converted the
Earth inerton field into a microwave electromagnetic radiation
[28]. In fact, the ancients possessed the detailed knowledge on
the constitution of the universe. For instance, we meet in {\it
the} $Bhagavad$-$g{\bar \imath}t{\bar a}$ [29]: "Yet there is
another nature, which is eternal and is transcendental to this
manifested and unmanifested matter. It is supreme and is never
annihilated. When all in this world is annihilated, that part
remains as it is."  Besides, quite recently Roy [30] has found the
clue to decoding the Vedic manuscripts and nowadays we could
ascribe another title to {\it the} {\it \d{R}gveda}, namely: {\it
Ultramodern Handbook on the Constitution of Space, Particle
Physics and Cosmology} (see also Ref. [28]). It is interesting to
note that the theory of space, which is developing by Bounias and
the author [16-19,22-25], {\it \textbf{exactly}} corresponds to
the pattern stated in the decoding {\it \d{R}gveda} [30].

The Earth's inerton waves are generated along West-East line owing
to the proper rotation of the Earth (in other words, the rotation
of the Earth results in its interaction with the space). In our
own experiment [31] (see also Ref. [28]) we could fix changes
caused by the Earth inerton field in samples that were stayed in a
resonator for 30 days. The construction of the resonator is very
simple; it consists of two faces bonded together in the top and
distant one from another at the bottom. The faces should be
oriented to the East and the West. Thus our resonator resembles a
small model pyramid in a certain sense. The result obtained
directly demonstrates the existence of the quantum mechanical
force on the macroscopic scale. In paper [31] we showed that an
external inerton field (the Earth's inerton field) influenced the
system studied in the same manner as the ultrasound. This allows
us to elaborate a facility that should record the quantum
mechanical force by using of rapid method. Figure 1 depicts the
stunning demonstration experiment: in the resonator the device
measures the inerton radiation of the Earth. The antenna of the
device is turned at a concrete frequency $\nu_{{\kern 1pt } 0}$,
which is generated by the proper electronic circuit. In the
presence of inerton radiation, frequency $\nu_{{\kern 1pt } 0}$
should be altered. If we turn the resonator on $90^\circ$ so that
its faces become oriented to the North and South, the device will
not record any radiation; along the North-South line the Earth
does not revolve and therefore in this direction no stable inerton
flows are available [31]. The device described is not a fantasy.
Similar devices constructed by engineers ("Demon," "Urga" and
"Tesey") have already been used by some geologists and architects
in Moscow and Kharkiv at the construction of model pyramids in
Russia and Ukraine.  The pyramids whose height varies from 10 to
44 m were built for both scientific studies and business purpose
(see Refs. [32,28]). The said devices measured noise that was
several times greater inside of the pyramid than that of outside.

\begin{figure}
\begin{center}
\includegraphics[scale=0.5]{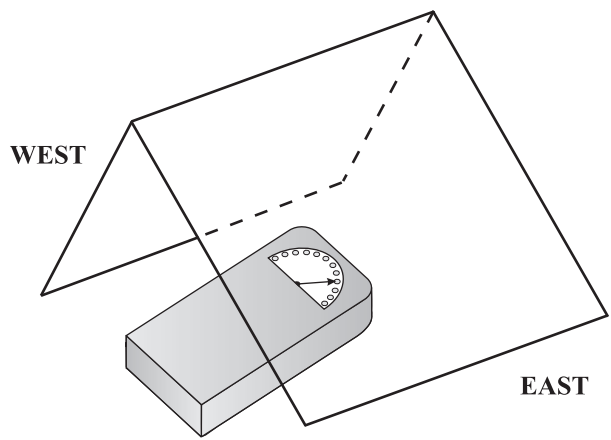}
\caption{Resonator made of two plates (for instance, transparent
organic glass) and the  device (rather pocket) that measures the
inerton radiation in the resonator interior.} \label{Figure1}
\end{center}
\end{figure}

\vspace{4mm}

\section{Concluding remarks}

\hspace*{\parindent} In this outline we have analysed basic
confusing points of conventional quantum mechanics. They are very
strong, though still are not virtually treated in the physics
literature. Nevertheless, we have to mention here the very
interesting recent work by Arunasalam [33] in which he considers
discrepancies in views on the fundamentals by famous physicists:
Einstein, Dirac, Feynman, Pauli, Bethe and others. Arunasalam
brings up the views on the fundamental problems expressed by
different sets of giants of physics and shows that the views are
in sharp contrast (for instance, covariance versus invariance,
relativistic versus nonrelativistic electron theories, etc.). He
is completely right when notes that "such conflicts put heavy
burden on the conscience of any physics teacher. After all, a good
physics teacher has a responsibility to explain to the students
which view is the correct one."

The problems posed by Arunasalam [33] indeed are considerably
importance. However, it seems that those problems arise from the
fact that the fundamental physics is still resting on an
undetermined basis. For instance, when we talk about charge
conjugation $\mathbf{C}$ and $\mathbf{CPT}$ violation, we neither
know nor understand what do the notions {\it charge}, {\it space},
and {\it time} mean exactly. These notions still are not
determined exactly at all! The same take place in all other
examples:  (i) the total angular momentum includes the vector
potential $\vec A$, but what does the latter mean?! (ii) The
problem "covariance versus invariance" is determined only in the
framework of phenomenological consideration; quantum mechanical
behaviour of the system studied change the problem very
significantly. (iii) The problem "relativistic and nonrelativistic
theories of the electron" strongly depends on the notions of spin
and the vector potential $\vec A$, which completely do not
understandable in the framework of conventional quantum theories.

Thus, only transition to sub atomic physics is capable to clarify
the discrepancies described by Arunasalam [33]. We need a detailed
microscopic theory of the real space, the notions of matter and
time derived from the space and the identification of the process
of motion of matter in the space. All these problems have been
raised by the author and partly already solved in works
[16-19,22-25,28]. The theoretical results obtained have found
rigorous experimental verification [31,34,35] (see also Ref. [28])
and the device (that has been developing) shown in Figure 1 is
reliable evidence of credibility of the proposed concept.

Further studies aimed at the understanding the constitution of the
real space and the generation of matter and physics laws in the
space promise a major breakthrough in fundamental science and
advance in technology.

\end{document}